\documentclass[conference]{IEEEtran}
\usepackage[noadjust]{cite}
\usepackage{breakurl}
\usepackage{xurl}
\ifCLASSINFOpdf
\else
\fi

\DeclareRobustCommand*{\IEEEauthorrefmark}[1]{%
  \raisebox{0pt}[0pt][0pt]{\textsuperscript{\footnotesize\ensuremath{#1}}}}
  \usepackage[misc]{ifsym}
\newcommand{\corrAuthor}{$^{\textrm{\Letter}}$}
\begin{document}
\bstctlcite{IEEEexample:BSTcontrol}

\title{The Case for Non-Volatile RAM in Cloud HPCaaS \vspace{-0.5ex}}

\author{
\IEEEauthorblockN{Yehonatan Fridman\IEEEauthorrefmark{1,2}, Re'em Harel\IEEEauthorrefmark{1,3,4} and
Gal Oren\IEEEauthorrefmark{3,5\hspace{0.1cm}$\corrAuthor$}}\\
\IEEEauthorblockA{\IEEEauthorrefmark{1}Department of Computer Science, Ben-Gurion University of the Negev, Israel}
\IEEEauthorblockA{\IEEEauthorrefmark{2}Israel Atomic Energy Commission}
\IEEEauthorblockA{\IEEEauthorrefmark{3}Scientific Computing Center, Nuclear Research Center – Negev, Israel}
\IEEEauthorblockA{\IEEEauthorrefmark{4}Department of Physics, Nuclear Research Center – Negev, Israel}
\IEEEauthorblockA{\IEEEauthorrefmark{5}Department of Computer Science, Technion – Israel Institute of Technology, Israel}
{\tt\small fridyeh@post.bgu.ac.il, reemha@bgu.ac.il, galoren@cs.technion.ac.il}
\vspace{-0.5ex}
}

\IEEEtitleabstractindextext{%
\begin{abstract}
HPC as a service (HPCaaS) is a new way to expose HPC resources via cloud services. However, continued effort to port large-scale tightly coupled applications with high interprocessor
communication to multiple (and many) nodes synchronously,
as in on-premise supercomputers, is still far from satisfactory due to network latencies. As a consequence, in said cases, HPCaaS is recommended to be used with one or few instances. In this paper we take the claim that new piece of memory hardware, namely Non-Volatile RAM (NVRAM), can allow such computations to scale up to an order of magnitude with marginalized penalty in comparison to RAM. Moreover, we suggest that the introduction of NVRAM to HPCaaS can be cost-effective to the users and the suppliers in numerous forms.

\end{abstract}

}
\maketitle

\IEEEdisplaynontitleabstractindextext

\IEEEpeerreviewmaketitle
\section{Supercomputing and the Cloud}

As the demand for more extensive calculations and simulations increases, huge amounts of data should be analyzed, manipulated, and stored. These needs ushered in a new Exascale era of stronger and more complex supercomputers \cite{heldens2020landscape, hoekstra2019multiscale, bernholdt2020survey}. This era was recently introduced with the first actual Exascale supercomputer – \textit{Frontier} \cite{schneider2022exascale}. \textit{Frontier} consists of more than 8 million compute cores that provide the peak computing power of 1.102 Exaflop/s.

Traditionally, said class of supercomputers can perform (almost) any kind of computation – from loosely to tightly coupled and from CFD simulations to deep-learning optimizations – with scalable performances compared to a single core or machine \cite{alexander2020exascale}. This scalability is mandatory for R\&D breakthroughs, as exact and more detailed calculations are the key to better science, and they tend to be latency-sensitive \cite{murphy2007effects}. Thus, the network architecture choice in modern supercomputers is of high speed, wide bandwidth, low latency, RDMA transmission enabled network, namely InfiniBand \cite{pfister2001introduction, grun2010introduction}, that is used to construct a complete system without (almost) any interference \cite{ajima2022optical, lu2022survey}. With switch latencies close to 100$ns$, and end-to-end latency of less than 1$\mu$s (end-point to end-point), InfiniBand interconnect achieves comparable performances to shared memory communication for high-performance and data-intensive applications. In scaling measurements of \textit{Summit} supercomputer, not long ago the world's fastest supercomputer, latency experiments from/to all 4,608 nodes showed a latency of 1.3$\mu$s for 0B messages and 3.54$\mu$s for 4KB messages with Mellanox EDR InfiniBand \cite{vergara2019scaling}.

Alongside the evolution of ‘classic’ supercomputing, a new architecture for HPC was raised – The Cloud, specifically HPC as a service (HPCaaS), which exposes HPC resources via cloud services \cite{parashar2013cloud}. The recent significant developments in the cloud allows better compute nodes than those in supercomputers, with denser computing power and much-improved access and elasticity \cite{aws2, aws3, gcp4}. As such, many small to low-medium scale calculations that usually needed access to supercomputers to be executed now can easily port to cloud services \cite{walkup2022best}. However, continued efforts to port large-scale tightly coupled applications with high interprocessor communication to multiple (and many) nodes synchronously, as in on-premise supercomputer, is still far from satisfactory and currently measured with a few hundred cores before scalability crashes \cite{fernandez2021evaluation}. This scalability issue results from the fabric (Ethernet, in this way or the other) with latency as high as one to two orders of magnitude than Infiniband on-premise supercomputers, even with dedicated optimizations \cite{fernandez2021evaluation, aws1, gcp1, gcp2, gcp3, phanekham2020measuring}. As a result, the cloud is usually recommended to be used in such scenarios with just one powerful instance (or a few), with as much computing power, memory and storage \cite{walkup2022best,aws6, zivanovic2016large}. For example, the new AWS Hpc6a.48xlarge contains 96 cores with 384GB of RAM \cite{aws2}, and the majority of cloud services already make available local SSD storage through SCSI and even NVMe interfaces \cite{kashyap2022nvme,guz2018performance,gupta2011evaluation, gcp5, aws4}. 

From the business perspective, the economic model of the public cloud, that much prefers elasticity over connectivity speed, can not support Infiniband networking efficiently and cost-effectively, let alone supplying a large amount of resources simultaneously, spontaneously and instantly \cite{netto2018hpc}. A unique exception that does not prove the rule is a supercomputer that was fully ported to the cloud: In Nov 2021, and for the first and only time since then, new supercomputers (mainly No. 10 on the TOP500 list \cite{top500}, the \textit{Voyager-EUS2}) were introduced entirely to one of the major cloud suppliers. As such, \textit{Voyager-EUS2} enjoys the two technologies’ benefits and disregards their weaknesses, except the most critical one in the cloud, the elasticity, that is bounded to the system resources.

In a seminal paper from 2009, Napper and Bientinesi wondered "Can Cloud Computing Reach The TOP500?" \cite{napper2009can}, and their concerns remain valid. Specifically, their observation that the flop/s obtained per dollar spent decrease exponentially with increasing computing cores and correspondingly, the cost for solving a linear system increases exponentially with the problem size -- is still very much in contrast to existing scalable supercomputers. "If cloud computing vendors are serious about targeting the HPC market," they stated, "different cost models must be explored. An obvious first step would be to offer better interconnects or nodes provisioned with more physical memory to overcome the slower network" \cite{napper2009can}. 

\section{Non-Volatile RAM in HPC Use cases}
Three primary resources of HPC (node-wise, excluding communication) are computing power, main memory and storage. Each component develops differently, which creates significant gaps that impair the scalability and performance of these systems. Large capacity of main memory is required to execute large problems, while its low latency is required to constantly feed the computing units with data. While computing power has increased exponentially over the years (mainly due to integrating heterogeneous hardware like CPUs and GPUs) \cite{denning2016exponential,top500}, DRAM scales slowly and presents no significant improvement in latency or bandwidth: The world's first DDR5 DRAM chip was officially launched in 2020 \cite{jing2020cache}, a decade after the DDR4 was introduced in 2011; and it has about the same latency as DDR4, just four times the density, and is only targeted to double the bandwidth. For example, in the flagship\textit{ Aurora} supercomputer \cite{aurora}, each compute node will hold the top density possible of 16 dies times 64GB per die, meaning $\sim$1TB. The DRAM is also expensive relative to capacity, creating a significant gap with the increasing demand for computational power.
Worse, the storage performances of HDDs and SSDs are extremely low in comparison to the DRAM, and exhibit no significant improvements in technology over the years. There are still 2-3 orders of magnitude between the performances of DRAM and storage devices \cite{luttgau2018survey}, and as such they create a huge bottleneck to scientific applications \cite{harrod2012journey}. However, DRAM is significantly more expensive than standard storage devices~\cite{luttgau2018survey} and its power consumption is higher, due to the cost of frequent refresh operations~\cite{fan2001memory}. Thus, building cost-effective, resilient, fast and power-efficient systems with large memory and storage spaces is a major challenge faced by the HPC community~\cite{oren2016memory,bergman2008exascale}.

A new building block in supercomputers is \textit{non-volatile RAM} (\textit{NVRAM}). Next-generation supercomputers are expected to integrate NVRAM devices. For example, \textit{Aurora}~\cite{Argonne1} is planned to integrate the emerging Intel Optane™ NVRAM. NVRAM provides better density and energy efficiency than DRAM
while providing DRAM-like performance: NVRAM comes in 128GB, 256GB and 512GB capacities (vastly larger than DRAM cards) \cite{intel1} and its read/write latencies are 400\% higher than those of DRAM, while the read/write bandwidths were 37\% and 8\% of those of DRAM, respectively \cite{hirofuchi2020prompt,hirofuchi2019preliminary}.
When configured in App-Direct mode these devices are byte-addressable and can be used by processes as NVRAM, while when configured in Memory or Flat mode these devices provide an extension to the volatile memory pool of the application, enabling the execution of much larger problems in a small number of compute nodes \cite{patil2021nvm}. 

Prior work on the use of NVRAM in HPC environments was focused on three main direct use-cases: 
(1) as fast storage for diagnostics~\cite{fridman2021assessing,hennecke2020daos,weiland2018exploiting}, 
(2) as fast persistence area for checkpointing~\cite{fridman2021assessing,ren2019easycrash}, and 
(3) as memory expansion to enable larger memory scientific workloads~\cite{fridman2021assessing,patil2019performance,weiland2019early,patil2021nvm}. 
The two former use-cases rely on the non-volatility of NVRAM for fast storage as a substitute for standard storage mediums~\cite{weiland2018exploiting,hennecke2020daos,fridman2021assessing}. 
The third use-case investigates the benefits of NVRAM in Memory mode, using its large capacity to enhance HPC capabilities~\cite{patil2019performance,patil2021nvm,fridman2021assessing}. 
In particular, a comprehensive investigation~\cite{fridman2021assessing} studies the main use-cases of Intel Optane™ NVRAM in HPC to improve scientific applications’ performances and fault tolerance.

\section{The Case for Non-Volatile RAM in HPCaaS}

It was shown that various HPC applications can benefit from scaling-in applications into fewer nodes by integrating NVRAM as memory expansion --- in terms of energy efficiency and performance \cite{patil2021nvm}. Patil et al. \cite{patil2021nvm} investigated the trade-off between performance degradation due to the higher latency of NVRAM and the savings in intra-node communication overheads when executing various scientific applications on a single NVRAM node (which contains DRAM and NVRAM) rather than on 4 DRAM-only nodes. This investigation assumed an Infiniband connection between the 4 DRAM-only nodes, and found that the shared memory buffer communication on the NVRAM node has up to half the latency and 3x the bandwidth than the Infiniband-connected DRAM-only nodes. Patil inferred that compute-bound applications, which utilize cache locality well, are able to run larger problem sizes on fewer compute nodes with a DRAM-NVRAM hybrid memory system, experiencing minimal performance degradation while providing significant cost and energy savings.

As a consequence of previous conclusions by Patil et al. \cite{patil2021nvm}, and given the Ethernet-based cloud connectivity, we hereby take the claim that NVRAM is an essential, and even crucial component in future cloud HPCaas instances, when considering memory and I/O bounded applications in general, or small to mid-range tightly-coupled computations in particular. We notice that even in current NVRAM capacity, it is possible -- with only a few factors of performance degradation in comparison to RAM \cite{fridman2021assessing, weiland2019early, xiang2022characterizing} -- to achieve up to an order of magnitude more byte-addressable memory, with much lower costs \cite{cost1, cost2}. In weak-scaled computations, that rely on RAM extensions, the presence of such memory can save not only the need to extend to other instances/nodes (and by that to reduce dramatic latency overheads), but also to significantly reduce the costs -- either for the clients (by requesting less resources and achieving more scalability) or the cloud supplier (with energy and scheduling savings). Furthermore, in applications that heavily rely on I/O, either for diagnostics or even for recoverability, the local NVRAM can serve as a reliable, vast and fast storage and memory combined, reducing the need to extend external distributed file systems over high latency communication.

\clearpage
\section*{Acknowledgment}
This work was supported by Pazy grant 226/20, ISF grant 1425/22, and the Lynn and William Frankel Center for Computer Science. Computational support was provided by the NegevHPC project~\cite{negevhpc}. 

\bibliographystyle{IEEEtran}
\bibliography{sample-base.bib}

\end{document}